\documentstyle[12pt]{article}
\date{}
\author{
Igor~V.~Beloussov and Vladimir~V.~Frolov  \\[5mm]
{\it Institute of Applied Physics, Academy of Sciences of Moldavia,\/} \\[3mm]
{\it Academy street~5, 277028 Kishinev, Moldavia\/}
}
\title{
Nonmonotonic Decay   \\
of Nonequilibrium Polariton Condensate   \\
in Direct-Gap Semiconductors
}
\begin{document}
\baselineskip=24pt
\maketitle
\begin{abstract}
\baselineskip=24pt

\ \ \ Time evolution of a nonequilibrium polariton condensate has
been studied in the framework of a microscopic approach. It has been
shown that due to polariton-polariton scattering a significant
condensate depletion takes place in a comparatively short time interval.
The condensate decay occurs in the form of multiple echo signals.
Distribution-function dynamics of noncondensate polaritons have
been investigated. It has been shown that at the initial stage of evolution
the distribution function has the form of a bell. Then oscillations arise in
the contour of the distribution function, which further transform into
small chaotic ripples. The appearance of a short-wavelength
wing of the distribution function has been demonstrated.
We have pointed out the enhancement and then partial extinction of the
sharp extra peak arising within the time interval characterized by small
values of polariton condensate density and its relatively slow changes.
\end{abstract}
%
%
\newpage

\section{Introduction\label{int_1.txt}}

In a great number of works concerning spatial and time evolution of laser
radiation in resonant media the classical description is used when the photon
and material fields are considered as interacting macroscopically occupied
coherent modes (condensates). When a quasiparticle quits a condensate mode,
that is usually taken into account by introduction of phenomenological
damping constants. In this case it is assumed that (1) the condensate decay
is exponential and (2) the condensate time life  is considerably longer than
the duration of the coherent nonlinear process being investigated.

From our point of view it is more consistent to use a microscopic approach
(see, e.g., \cite{Elesin,Sc_1988,Keldysh_3}),
where the dynamics of interacting condensate modes and quantum fluctuations
appearing in a system as a result of real scattering processes of
quasiparticles are described on equal footing. In other words, one
should take into account (1) the interaction between condensates,
(2) excitation of noncondensate particles, and
(3) the backward influence of non-condensate particles on the condensates.

The microscopic  approach is used in this work to describe the
of time evolution of the system of initially coherent high-density
excitons and photons.

It is known \cite{1} that  coherent
electromagnetic radiation resonant to the isolated exciton
energy level excites in the crystal the coherent polariton wave
with the wave vector  ${\bf k}_{0} \neq 0$  ---
the nonequilibrium polariton condensate.  Different scattering
processes accompanying its propagation lead to
the loss of initial coherence of the polariton wave,
complete  or partial depletion of condensate, excitation of
polaritons characterized by some statistics with wave vector
${\bf k} \neq {\bf k}_{0} $,
and other phenomena.

At sufficiently high excitation energies the processes of polariton
scattering due to exciton-exciton collisions prevail.
This scattering mechanism is of considerable interest
due to the recent experimental investigations \cite{2}
and  many interesting physical results obtained in theoretical study of
dynamic and kinetic processes in the system of interacting polaritons.

      According to \cite{4,5} exciton-exciton
scattering is of significant importance in the situation where
coherent polaritons are excited in a certain spectral region
where energy and momentum conservation laws allow real
processes of two-quantum  excitation of polaritons from the
condensate. These processes lead to instability of the
completely condensed state of the polariton system.
Existence of this spectral region situated around the
isolated exciton resonance is due to the peculiarities of the
polariton  dispersion law.

     In \cite{4,5} the energy spectrum of non-condensate
polaritons, arising as a result of decay of coherent polariton
wave, is studied.  According to \cite{4,5} in some regions
of {\bf k} space the energy spectrum does not exist.

     It should be mentioned that investigations performed in \cite{4,5}
are based on the model for\-mally ana\-lo\-gous to that used by
Bogoliubov in \cite{6} to stu\-dy the equi\-lib\-rium sys\-tem
of weak\-ly nonideal Bose gas. In the nonequilibrium situation
considered in \cite{4,5}, when decay of the polariton condensate
and excitation of noncondensate polaritons take place,
this model is adequate to the real situation only
at the initial stage of the condensate decay  when the number of
polaritons in the condensate  is still  much greater than the total number
of non-condensate polaritons. But this stage is essentially
{\sl nonstationary\/} and the occurrence of condensate instability
provides evidence for that. The study of the energy spectrum implies the
determination of the {\sl steady states\/} of the system  \cite{7}.
That is why the results of \cite{4,5}, concerning the energy
spectrum of the system, based on the above mentioned model,
in our opinion cannot be regarded as well grounded. This remark
refers to the relatively recent works  \cite{m_m_sh,b_m_r_sh} as well.

     Because of essential nonstationarity  of the processes
in the system  methods of  nonequilibrium mechanics should
be used to describe it adequately. Derivation of the equations that
describe  kinetics  of the polariton condensate decay and
excitation of quantum fluctuations has some specific features owing
to degeneracy in the system. As the energy and resulting wave
vector of two non-condensate polaritons can be equal to the energy
and wave vector of two condensate polaritons,  respectively,
there is degeneracy of two-particle states. Moreover, the presence
of the condensate in the system also leads to degeneracy  due to
its macroscopic amplitude \cite{8}.

     The correct description of the system with degeneracy requires
the introduction of abnormal distribution functions \cite{9}
$$
 \Psi_{{ \bf k}_0} = \langle \Phi_{{\bf k}_0 } \rangle
$$
and
$$
 F_{{\bf k}} = \langle \Phi_{{\bf k}} \Phi_{2{\bf k}_0
- {\bf k}} \rangle
$$
together  with the normal (usual) ones
$$
N_{\mbox{\small \bf k }} =
\langle \Phi_{{\bf k}}^+ \Phi_{{\bf k}} \rangle.
$$
Here $\Phi_{{\bf k}}^+ (\Phi_{{\bf k}})$
are Bose operators of creation (annihilation) of a polariton on
the lower branch with the wave vector   {\bf k}. The appearance
of abnormal averages
indicates a breaking of the selection rules connected with the
gauge invariance of the system \cite{9,10,11}.
In this problem the breaking of gauge symmetry arises as a
result of the noninvariant initial condition due to the assumption that there is
a condensate in the system at the initial moment $t = t_{0}$.

     An earlier attempt to obtain kinetic equations for polaritons
excited in semiconductors by the external classic field was
made in \cite{12,13}.  However, in \cite{12,13}
degeneracy of two-particle states  was not taken into account
and the abnormal distribution functions  $F_{{\bf k}}$
were not introduced. So one should expect the equations obtained
in \cite{12,13} to possess unphysical singularity.

    Kinetic equations describing evolution of partially coherent
polaritons that take into account the degeneracy were obtained in
\cite{14,15} using the nonequilibrium Green's-function technique
\cite{16,kad_b}, presented by the authors in terms of functionals. They
coincide  with the equations obtained  in \cite{17}
using the method of nonequilibrium statistical
operator \cite{18}, and do not possess unphysical singularities.

     According to \cite{15,17} the kinetics of partially
coherent polaritons is described in the Born approximation by
the closed set of non\-linear integ\-ro\-dif\-feren\-tial equa\-tions
for the coherent part of polariton field $\Psi_{{\bf k}_0} $
and
the normal
$$
n_{\bf k} = N_{\bf k} -
\delta_{{\bf k},{\bf k}_0} |\Psi_{{\bf k}_0}|^2
$$
and abnormal
$$
 f_{\bf k} =  F_{\bf k} -
\delta_{{\bf k},{\bf k}_0} \Psi_{{\bf k}_0}^2
$$
distribution functions.
In the absence of quantum fluctuations, described by the functions
$n_{\bf k}$       and       $ f_{\bf k}$,
the equations for them
become identities, and the equation for
$\Psi_{{\bf k}\alpha} $
($\alpha $    is the number of polariton branch) coincides with that
obtained in \cite{22} for the system of interacting coherent
excitons and photons.
In another particular case, when
$\Psi_{{\bf k}_0} = 0$
and
$f_{\bf k} = 0$, the equations obtained in \cite{15,17}
are reduced to
the usual kinetic equation for the distribution function
$N_{\bf k}$       (see, e.g., \cite{18,23}).

The right-hand sides of kinetic equations obtained in \cite{15,17}
include terms linear in the constant of exciton-exciton interaction
${\nu}>0$ and the ones $\sim\!{\nu}^{\,2}$. Terms $\sim\!\nu$ correspond to
the self-consistent field approximation, which neglects the higher-order
correlation functions. In this approximation the processes of two-particle
excitations of polaritons from the condensate, backward processes, and
transformation of a created pair of noncondensate polaritons into
another pair with the same value of total momentum are taken into account.
It describes the fastest processes in the system and
is sufficient for study of the early (before-kinetic) stage of the system's
evolution.

Terms $\sim\!{\nu}^{\,2}$ take into account scattering processes in which only
one polariton belongs to the condensate. Therefore, they differ from
zero only if noncondensate polaritons exist in the system. Terms
$\sim\!{\nu}^{\,2}$ in comparison with terms $\sim\!{\nu}$ describe slower
changes of the system characteristics and are significant only at the kinetic
stage of evolution.

Our study of time evolution of the nonequilibrium polariton system is
based on the self-consistent field approximation. This approximation
has shown a good performance in the theoretical study of states of
the electron-hole subsystem in semiconductors  that appear just after the
transmission of the  front of the ultrashort laser pulse
 \cite{Sc_1986,Sc_1988,Comte}. The part of evolution equations obtained in
our work for the stationary case are in many respects similar to the equations
given in \cite{Sc_1986,Sc_1988,Comte}. Nevertheless, that slight distinctions
are important and lead to significant physical consequencies.

The stationary equations obtained in \cite{Sc_1986,Sc_1988,Comte}
possess nontrivial solutions describing states of the
electron-hole subsystem. The decay of these states occurs only due to
incoherent relaxation processes. In \cite{Sc_1986,Sc_1988,Comte} it is
assumed that these processes do not have time to happen
during the period of pulse action. This assumption justifies the use
of the self-consistent field approximation.
The attempt to find the steady-state solution of similar equations
in the framework of the physical problem of this paper leads to instabilities.
Their physical nature has been discussed above. As noted, the
appearance of the instability points leads to  essential non-stationarity
of the processes in the system and requires the return to the
starting (nonstationary) equations.

Moreover, the excitation of the great number of noncondensate  modes can
lead to significant depletion of the condensate. So to take into account
this phenomenon it is necessary, along with equations similar to equations
in \cite{Sc_1986,Sc_1988,Comte} to consider
an additional equation for the condensate wave function $\Psi_{{\bf k}_0}$.
Note that in \cite{Sc_1986,Sc_1988,Comte} backward influence of the
electron-hole subsystem on the laser radiation field was not taken into
account.

Note also \cite{Bob_Mosk,Bob_Mosk_Sh,Keldysh_1,Keldysh_2,Bob_Mosk_Cam}
where the self-consistent field approximation allowed us to take into account
the biexciton complex structure.
%
%

\section{System Hamiltonian\label{ham_2.txt}}

   We start from the Hamiltonian
\begin{equation}
{\hat H} = {\hat H}_{0} + {\hat H}_{\mbox{\it \small int}}\>,
\label{eq_1}
\end{equation}
\smallskip
\begin{equation}
{\hat H}_{0} = {\sum_{\bf k}}{\sum_{{\alpha}_{1},{\alpha}_{2} = 1,2}}
{(h_{\bf k})_{{\alpha}_{1}{\alpha}_{2}}}{{\hat \varphi}_{{\bf k}{\alpha}_{1}}
^{\>+}}{{\hat \varphi}_{{\bf k}{\alpha}_{2}}}\>,\qquad
h_{\bf k} = \hbar{{\omega}_{\bf k}^{ \perp}}\left(\frac{1 +
{\tau}_{3}}{2}\right)
+ \hbar{{\omega}_{\bf k}}\left(\frac{1 - {\tau}_{3}}{2}\right)
+ {\eta}_{\bf k}{\tau}_{2}\,,
\label{eq_2}
\end{equation}
\smallskip
\begin{eqnarray}
{\hat H}_{\mbox{\it \small int}}  & = & \frac{1}{2V}{\sum_{{\bf k}_{1},
\ldots , {\bf k}_{4}}}\>\sum_{{\alpha}_{1},\ldots , {\alpha}_{4} = 1,2}
{\nu}_{\,{\bf k}_{1} - {\bf k}_{4}}{\,\delta}_{{\bf k}_{1} + {\bf k}_{2},
{\bf k}_{3} + {\bf k}_{4}} \nonumber\\
&  \times  & \left({\frac{1 + {\tau}_{3}}{2}}\right)_{{\alpha}_{1},{\alpha}_{3}}
\left({\frac{1 + {\tau}_{3}}{2}}\right)_{{\alpha}_{2},{\alpha}_{4}}
{{\hat \varphi}_{{{\bf k}_{1}}{\alpha}_{1}}^{\>+}}
{{\hat \varphi}_{{{\bf k}_{2}}{\alpha}_{2}}^{\>+}}
{{\hat \varphi}_{{{\bf k}_{3}}{\alpha}_{3}}}
{{\hat \varphi}_{{{\bf k}_{4}}{\alpha}_{4}}}\>. \label{eq_3}
\end{eqnarray}

   In~(\ref{eq_1})-(\ref{eq_3})  the  following  notations are used.
The operators ${\hat \varphi}_{{\bf k} \alpha }$
(${\hat \varphi}_{{\bf k} \alpha }^{\> + }$)  of  exciton
(when $\alpha  = 1$) or photon (when $\alpha    = 2$)
annihilation  (creation)  in  the state with wave vector ${\bf k}$
obey  Bose-type commutation relations:
$$
\left[
{{\hat \varphi}_{{{\bf k}_{1}}{\alpha}_{1}}},
{{\hat \varphi}_{{{\bf k}_{2}}{\alpha}_{2}}^{\>+}}\right] =
{\delta}_{{\bf k}_{1},{\bf k}_{2}}{\delta}_{{\alpha}_{1},{\alpha}_{2}}\,,
\qquad
\left[
{{\hat \varphi}_{{{\bf k}_{1}}{\alpha}_{1}}},
{{\hat \varphi}_{{{\bf k}_{2}}{\alpha}_{2}}}\right] =
\left[{{\hat \varphi}_{{{\bf k}_{1}}{\alpha}_{1}}^{\>+}},
{{\hat \varphi}_{{{\bf k}_{2}}{\alpha}_{2}}^{\>+}}\right] = 0\>.
$$
The  Pauli  matrices ${\tau}_{1}$, ${\tau}_{2}$,  and  ${\tau}_{3}$
are taken in the standard representation \cite{16}.
The frequencies
${\omega}_{\bf k}$
and
${\omega}_{\bf k}^{\,\perp}$
are given by the expressions
${\omega}_{\bf k} = c\left|{\bf k}\right|{\epsilon}_{B}^{-1/2}\>,
{\omega}_{\bf k}^{\,\perp} = {\omega}^{\,\perp} + \hbar{\bf k}^{2}/2m\>,$
where $c$  is the vacuum velocity  of light,
$\hbar{\omega}^{\,\perp}$ is the exciton formation energy in
the  band
$\hbar {\omega}_{\bf k}^{\,\perp}$, and $m$ is its  effective mass.
The background dielectric  function ${\epsilon}_{B}$  includes  the
contributions  from all excitations  in  a  crystal except the
excitons of the isolated band $\hbar {\omega}_{\bf k}^{\,\perp}$.
In  the  vicinity  of  the  exciton resonance
${\omega}_{\bf k} \approx {\omega}_{\bf k}^{\,\perp}$ weak
frequency  dependence of ${\epsilon}_{B}$ can be neglected.

     The Hamiltonian~(\ref{eq_1}) describes the system of interacting
dipole-active excitons and photons with transverse polarization in
an infinite crystal ($V \rightarrow \infty$, $V$  is the
quantization volume) in the vicinity of an isolated exciton resonance
${\omega}_{\bf k} \approx {\omega}_{\bf k}^{\,\perp}$. In other
words, we assume that the relations
$\hbar\left|{\omega}_{\bf k} - {\omega}_{\bf k}^{\,\perp}\right| \ll \hbar
{\omega}_{\bf k}^{\,\perp}\>,
{\Delta}{\cal E}$  are imposed on the photon frequency
$\hbar {\omega}_{\bf k}$, the energy of the exciton
$\hbar {\omega}_{\bf k}^{\,\perp}$,  and the minimum energy gap
${\Delta}{\cal E}$  between the exciton
band $\hbar {\omega}_{\bf k}^{\,\perp}$  and any other one.
Besides that we suppose that the constants  of  exciton-photon
(${\eta}_{\bf k}={\eta}_{-\bf k}$) and exciton-exciton
(${\nu}_{\bf k}={\nu}_{-\bf k}$)  interactions  are  small  enough:
$0 <  {\eta}_{\bf k}/\hbar \ll {\omega}_{\bf k}^{\,\perp}\>,
\quad 0 < ({\nu}_{\bf k}/\hbar){\bar n} \ll {\omega}_{\bf k}^{\,\perp}
$ ($\bar n$ is an average density of excitons in the system).
These  assumptions allow us to retain in~(\ref {eq_2})
only  resonant  terms  and  treat the exciton-exciton
interaction as a small perturbation.

    The  quadratic  part~(\ref{eq_2}) of  Hamiltonian~(\ref{eq_1})
can be reduced to the diagonal form
\begin{equation}
{\hat H}_{0}\,=\,{\sum_{\bf k}}{\sum_{\alpha\,=\,1,2}}
\hbar{{\Omega}_{{\bf k}\alpha}}
{\hat \Phi}_{{\bf k} \alpha }^{\>+}{\hat \Phi}_{{\bf k} \alpha }
\label{eq_5}
\end{equation}
by  transition  to  the polariton Bose operators
${\hat \Phi}_{{\bf k} \alpha}$     and ${\hat \Phi}_{{\bf k} \alpha }^{\>+}$
using the unitary transformation
$$
{{\hat \varphi}_{{{\bf k}}{\alpha}_{1}}} =
{\sum_{{\alpha}_{2} = 1,2}}
{(U_{\bf k})_{{\alpha}_{1}{\alpha}_{2}}}
{{\hat \Phi}_{{\bf k}{\alpha}_{2}}}\,,\qquad
U_{\bf k} = \frac{1 - i{\tau}_{1}L_{\bf k}}{\sqrt{1 + L_{\bf k}^{2}}}\>.
$$
$L_{\bf k}$ is    a    function    determined    by   the   equation
${\eta}_{\bf k}L_{\bf k}^{2} + \hbar\left({\omega}_{\bf k} -
{\omega}_{\bf k}^{\,\perp}\right)L_{\bf k} - {\eta}_{\bf k} = 0$.
In the polariton representation we have
\begin{eqnarray}
{\hat H}_{\mbox{\it \small int}}  & = & \frac{1}{2V}{\sum_{{\bf k}_{1},
\ldots , {\bf k}_{4}}}\>\sum_{{\alpha}_{1},\ldots , {\alpha}_{4} = 1,2}
{\nu}_{\,{\bf k}_{1} - {\bf k}_{4}}{\,\delta}_{{\bf k}_{1} + {\bf k}_{2},
{\bf k}_{3} + {\bf k}_{4}} \nonumber\\
&  \times  &
\left(
{\cal P}_{  {\bf k}_{1} {\bf k}_{3} }\right)_{{\alpha}_{1} {\alpha}_{3}}
\left(
{\cal P}_{  {\bf k}_{2} {\bf k}_{4} }\right)_{{\alpha}_{2} {\alpha}_{4}}
{{\hat \Phi}_{{{\bf k}_{1}}{\alpha}_{1}}^{\>+}}
{{\hat \Phi}_{{{\bf k}_{2}}{\alpha}_{2}}^{\>+}}
{{\hat \Phi}_{{{\bf k}_{3}}{\alpha}_{3}}}
{{\hat \Phi}_{{{\bf k}_{4}}{\alpha}_{4}}}\>,  \label{eq_6}
\end{eqnarray}
where
$$
{\cal P}_{  {\bf k}_{1} {\bf k}_{2}  } =
{\cal P}_{  {\bf k}_{2} {\bf k}_{1}  }^{\>+} =
U_{{\bf k}_{1}}^{\>+}
\left(\frac{1 + {\tau}_{3}}{2}\right)
U_{{\bf k}_{2}}\>.
$$

    Hamiltonian~(\ref{eq_1})  is  invariant with respect to the gauge
transformation $\hat R\,=\,\exp \left(i\gamma \hat N\right)$
where $\gamma$  is an arbitrary real parameter,
$\hat N\,=\,{\sum_{\bf k}}{\sum_{\alpha\,=\,1,2}}
{{\hat \Phi}_{{\bf k}{\alpha}}^{\>+}}{{\hat \Phi}_{{\bf k}{\alpha}}}
$  is the operator of the total number of polaritons  in the system.
%
%

\section{Self-consistent field approximation\label{appr_3}}

     In the Heisenberg picture operators ${\check \Phi}_{{\bf k}{\alpha}}(t)$
are governed by the
equation of motion
\begin{equation}
i\hbar \frac{d}{dt} {\check \Phi}_{{\bf k}{\alpha}}(t) = \left[
{\check \Phi}_{{\bf k}{\alpha}}(t),\check H \right]   \label{eq_60}
\end{equation}
and average value of a dynamic quantity $A$ is given by
${\langle A \rangle}_{t} = \mbox{\rm Tr}\, {\check \rho}{\check A}(t)$.
Here $\check \rho$ is the density matrix, which describes the polariton
distribution at the initial moment of time $t = t_{0}$.

    The presence of the condensate in the system implies that the coherent part
of the polariton field
\begin{equation}
{\Psi}_{{\bf k}{\alpha}}(t) = {\langle
{\Phi}_{{\bf k}{\alpha}}
\rangle}_{t}  \label{eq_61}
\end{equation}
is nonzero. We write down the coherent part explicitly
\begin{equation}
{\check \Phi}_{{\bf k}{\alpha}}(t) = {\Psi}_{{\bf k}{\alpha}}(t) +
{\check \chi }_{{\bf k}{\alpha}}(t)\>.  \label{eq_62}
\end{equation}
According to definition~(\ref{eq_61}), we have
${\langle {\chi}_{{\bf k} \alpha } \rangle }_{t} = 0$.

      Using~(\ref{eq_62}) we present the Hamiltonian
$\check H = {\check H}_{0} + {\check H}_{\mbox{\it \small int}}$ in the form
$\check H = {\check H}_{1}(t) + {\check H}_{2}(t)$, where the operator
${\check H}_{1}(t)$  includes only linear and quadratic terms with
respect  to  ${\check \chi }_{{\bf k} \alpha }(t)$  and
${\check \chi }_{{\bf k} \alpha }^{\> +}(t)$.
The operator ${\check H}_{2}(t)$  includes products of three and four
operators ${\check \chi }_{{\bf k} \alpha }(t)$  and
${\check \chi }_{{\bf k} \alpha }^{\> +}(t)$.

    Further  we shall treat ${\check H}_{2}(t)$ in  the  self-consistent field
approximation. For this  purpose we make the formal substitution \cite{Comte}
$$
{\check \chi }_{  {\bf k}_{1} {\alpha}_{1}       }^{\>+}(t)
{\check \chi }_{  {\bf k}_{2} {\alpha}_{2}       }^{\>+}(t)
{\check \chi }_{  {\bf k}_{3} {\alpha}_{3}       }      (t)
{\check \chi }_{  {\bf k}_{4} {\alpha}_{4}       }      (t)
\longrightarrow
$$
$$
{\check \chi }_{  {\bf k}_{1} {\alpha}_{1}       }^{\>+}(t)
{\check \chi }_{  {\bf k}_{2} {\alpha}_{2}       }^{\>+}(t)
\langle
{\chi }_{   {\bf k}_{3}  {\alpha}_{3} }
{\chi }_{   {\bf k}_{4}  {\alpha}_{4} }
\rangle _{t}
  +
{\check \chi }_{    {\bf k}_{3} {\alpha}_{3}    }  (t)
{\check \chi }_{    {\bf k}_{4} {\alpha}_{4}    }  (t)
\langle
{\chi }_{   {\bf k}_{1}  {\alpha}_{1}     }^{\> +}
{\chi }_{   {\bf k}_{2}  {\alpha}_{2}     }^{\> +}
\rangle _{t}
  +
{\check \chi }_{    {\bf k}_{1}  {\alpha}_{1}    }^{\>+}(t)
{\check \chi }_{    {\bf k}_{3}  {\alpha}_{3}    }      (t)
\langle
{\chi }_{   {\bf k}_{2}  {\alpha}_{2}   }^{\> +}
{\chi }_{   {\bf k}_{4}  {\alpha}_{4}   }
\rangle _{t}
$$
$$
 +
{\check \chi }_{  {\bf k}_{1}  {\alpha}_{1}  }^{\>+}(t)
{\check \chi }_{  {\bf k}_{4}  {\alpha}_{4}  }      (t)
\langle
{\chi }_{   {\bf k}_{2}  {\alpha}_{2}   }^{\> +}
{\chi }_{   {\bf k}_{3}  {\alpha}_{3}   }
\rangle _{t}
  +
{\check \chi }_{   {\bf k}_{2}  {\alpha}_{2}   }^{\>+}(t)
{\check \chi }_{   {\bf k}_{3}  {\alpha}_{3}   }      (t)
\langle
{\chi }_{  {\bf k}_{1}  {\alpha}_{1}   }^{\> +}
{\chi }_{  {\bf k}_{4}  {\alpha}_{4}   }
\rangle _{t}
  +
{\check \chi }_{  {\bf k}_{2}  {\alpha}_{2}  }^{\>+}(t)
{\check \chi }_{  {\bf k}_{4}  {\alpha}_{4}  }      (t)
\langle
{\chi }_{   {\bf k}_{1}  {\alpha}_{1}   }^{\> +}
{\chi }_{   {\bf k}_{3}  {\alpha}_{3}   }
\rangle _{t} \>,
$$
\smallskip
$$
{\check \chi }_{   {\bf k}_{2}  {\alpha}_{2}   }^{\>+}(t)
{\check \chi }_{   {\bf k}_{3}  {\alpha}_{3}   }      (t)
{\check \chi }_{   {\bf k}_{4}  {\alpha}_{4}   }      (t)
\longrightarrow
$$
$$
{\check \chi }_{   {\bf k}_{2}  {\alpha}_{2}   }^{\>+}(t)
\langle
{\chi }_{    {\bf k}_{3}  {\alpha}_{3}    }
{\chi }_{    {\bf k}_{4}  {\alpha}_{4}     }
\rangle_{t}
  +
{\check \chi }_{   {\bf k}_{3}  {\alpha}_{3}   }(t)
\langle
{\chi }_{   {\bf k}_{2}  {\alpha}_{2}   }^{\> +}
{\chi }_{   {\bf k}_{4}  {\alpha}_{4}   }
\rangle _{t}
 +
{\check \chi }_{   {\bf k}_{4}  {\alpha}_{4}   }(t)
\langle
{\chi }_{   {\bf k}_{2}  {\alpha}_{2}   }^{\> +}
{\chi }_{   {\bf k}_{3}  {\alpha}_{3}   }
\rangle _{t} \>.
$$
As a result, ${\check H}_{2}(t)$  (and  $\check H$) has the same
operator structure as ${\check H}_{1}(t)$. Performing transformation from
the operators ${\check \chi }_{{\bf k}{\alpha}}(t)$ back to the operators
${\check \Phi }_{{\bf k}{\alpha}}(t)$ we obtain
\newpage
$$
\check H   =   E_{0}(t)  +  {\sum_{\bf k}}{\sum_{\alpha = 1,2}}
\hbar{{\Omega}_{{\bf k}\alpha}}
{{\check \Phi}_{{\bf k}{\alpha}}^{\>+}}(t){{\check \Phi}_{{\bf k}{\alpha}}}(t)
   +   \frac{1}{2V}{\sum_{{\bf k}_{1},
\ldots , {\bf k}_{4}}}\>\sum_{{\alpha}_{1},\ldots , {\alpha}_{4} = 1,2}
{\delta}_{{\bf k}_{1} + {\bf k}_{2},{\bf k}_{3} + {\bf k}_{4}}
$$
$$
\times
\left\{
\left[{\nu}_{\,{\bf k}_{1} - {\bf k}_{4}}
\left(
{\cal P}_{  {\bf k}_{1} {\bf k}_{3}  }\right)_{{\alpha}_{1} {\alpha}_{3}}
\left(
{\cal P}_{ {\bf k}_{2}  {\bf k}_{4}  }\right)_{{\alpha}_{2} {\alpha}_{4}}
  +
{\nu}_{\,{\bf k}_{1} - {\bf k}_{3}}
\left(
{\cal P}_{  {\bf k}_{1} {\bf k}_{4}  }\right)_{{\alpha}_{1} {\alpha}_{4}}
\left(
{\cal P}_{  {\bf k}_{2} {\bf k}_{3}  }\right)_{{\alpha}_{2} {\alpha}_{3}}
\right] \right.
$$
$$
\times \biggl[
{\check \Phi }_{  {\bf k}_{1}  {\alpha}_{1}   }^{\>+}(t)
{\check \Phi }_{  {\bf k}_{3}  {\alpha}_{3}   }      (t)
\langle
{\Phi }_{        {\bf k}_{2}  {\alpha}_{2}   }^{\>+}
{\Phi }_{        {\bf k}_{4}  {\alpha}_{4}   }
\rangle _{t}
 +
\frac{1}{2}
{\check \Phi }_{  {\bf k}_{1}  {\alpha}_{1}   }^{\>+}(t)
{\check \Phi }_{  {\bf k}_{2}  {\alpha}_{2}   }^{\>+}(t)
{\langle
{\Phi }_{        {\bf k}_{3}  {\alpha}_{3}   }
{\Phi }_{        {\bf k}_{4}  {\alpha}_{4}   }
\rangle}_{t}
$$
\begin{equation}
-
\left.
\left.
2{\check \Phi }_{  {\bf k}_{1}  {\alpha}_{1}  }^{\>+}(t)
{\Psi }_{   {\bf k}_{2}  {\alpha}_{2}  }^{\>\ast}(t)
{\Psi }_{  {\bf k}_{3}  {\alpha}_{3}  }      (t)
{\Psi }_{  {\bf k}_{4}  {\alpha}_{4}  }      (t)
\right]
  +   \mbox{\rm H.c.}
\right\} \>,  \label{eq_63}
\end{equation}
where  $E_{0}(t)$  is a $c$-number function.

      With the help of  Hamiltonian~(\ref{eq_63}) and equations of
motion~(\ref{eq_60}) we obtain the set of equations, which describes time
evolution of the system of partially coherent polaritons:
$$
\left[ i\hbar \frac{d}{dt} - \hbar {\Omega}_{{\bf k}_{1} {\alpha}_{1}}\right]
{\Psi }_{  {\bf k}_{1}  {\alpha}_{1}  }      (t) =
\frac{1}{V}{\sum_{{\bf k}_{2}, {\bf k}_{3}, {\bf k}_{4}}}\>
\sum_{{\alpha}_{2}, {\alpha}_{3}, {\alpha}_{4} = 1,2}
{ \delta}_{{\bf k}_{1} + {\bf k}_{2},{\bf k}_{3} + {\bf k}_{4}}
$$
$$
\times
\left[{\nu}_{\,{\bf k}_{1} - {\bf k}_{4}}
\left(
{\cal P}_{ {\bf k}_{1} {\bf k}_{3} }\right)_{{\alpha}_{1} {\alpha}_{3}}
\left(
{\cal P}_{  {\bf k}_{2} {\bf k}_{4} }\right)_{{\alpha}_{2} {\alpha}_{4}}
  +
{\nu}_{\,{\bf k}_{1} - {\bf k}_{3}}
\left(
{\cal P}_{  {\bf k}_{1} {\bf k}_{4} }\right)_{{\alpha}_{1} {\alpha}_{4}}
\left(
{\cal P}_{ {\bf k}_{2} {\bf k}_{3}  }\right)_{{\alpha}_{2} {\alpha}_{3}}
\right]
$$
\begin{equation}
\times
\left[
\frac{1}{2}
{\Psi }_{   {\bf k}_{2}  {\alpha}_{2}  }^{\>\ast}(t)
{\Psi }_{  {\bf k}_{3}  {\alpha}_{3}  }      (t)
{\Psi }_{  {\bf k}_{4}  {\alpha}_{4}  }      (t)
+
n({\bf k}_{2}, {\alpha}_{2}; {\bf k}_{3}, {\alpha}_{3}|t)
{\Psi }_{  {\bf k}_{4}  {\alpha}_{4}  }      (t)
+
\frac{1}{2}
{\Psi }_{   {\bf k}_{2}  {\alpha}_{2}  }^{\>\ast}(t)
f({\bf k}_{3}, {\alpha}_{3}; {\bf k}_{4}, {\alpha}_{4}|t)
\right]\>;    \label{eq_64}
\end{equation}
\bigskip
$$
\left[
i\hbar \frac{d}{dt} +
\hbar
\left(
{\Omega}_{{\bf k}_{1}{\alpha}_{1}} -
{\Omega}_{{\bf k}_{2}{\alpha}_{2}}
\right)
\right]
n({\bf k}_{1}, {\alpha}_{1}; {\bf k}_{2}, {\alpha}_{2}|t)
=
-
\frac{1}{V}{\sum_{{\bf k}_{3}, {\bf k}_{4}, {\bf k}_{5}}}\>
\sum_{{\alpha}_{3}, {\alpha}_{4}, {\alpha}_{5} = 1,2}
{ \delta}_{{\bf k}_{1} + {\bf k}_{3},{\bf k}_{4} + {\bf k}_{5}}
$$
$$
\times
\left[{\nu}_{\,{\bf k}_{1} - {\bf k}_{5}}
\left(
{\cal P}_{  {\bf k}_{4} {\bf k}_{1}  }\right)_{{\alpha}_{4}{\alpha}_{1}}
\left(
{\cal P}_{  {\bf k}_{5} {\bf k}_{3}  }\right)_{{\alpha}_{5}{\alpha}_{3}}
  +
{\nu}_{\,{\bf k}_{1} - {\bf k}_{4}}
\left(
{\cal P}_{  {\bf k}_{5} {\bf k}_{1}  }\right)_{{\alpha}_{5} {\alpha}_{1}}
\left(
{\cal P}_{  {\bf k}_{4} {\bf k}_{3}  }\right)_{{\alpha}_{4} {\alpha}_{3}}
\right]
$$
$$
\times
\biggl\{
n^{\>*}({\bf k}_{2}, {\alpha}_{2}; {\bf k}_{5}, {\alpha}_{5}|t)
\left[
n^{\>\ast}({\bf k}_{3}, {\alpha}_{3}; {\bf k}_{4}, {\alpha}_{4}|t)
 +
{\Psi }_{  {\bf k}_{3}  {\alpha}_{3}  }      (t)
{\Psi }_{   {\bf k}_{4}  {\alpha}_{4}  }^{\>\ast}(t)
\right]
$$
$$
+
\frac{1}{2}
f({\bf k}_{2}, {\alpha}_{2}; {\bf k}_{3}, {\alpha}_{3}|t)
\left[
f^{\>\ast}({\bf k}_{5}, {\alpha}_{5}; {\bf k}_{4}, {\alpha}_{4}|t)
 +
{\Psi }_{  {\bf k}_{5}  {\alpha}_{5}  }^{\>\ast}(t)
{\Psi }_{   {\bf k}_{4}  {\alpha}_{4}  }^{\>\ast}(t)
\right]
\biggr\}
$$
\begin{equation}
 -
\left\{
\mbox{\rm Idem}
\left[
\left(
{\bf k}_{1}{\alpha}_{1}
\right)
\leftrightarrow
\left(
{\bf k}_{2}{\alpha}_{2}
\right)
\right]
\right\}^{\>\ast}\>;  \label{eq_65}
\end{equation}
\bigskip
$$
\left[
i\hbar \frac{d}{dt} -
\hbar
\left(
{\Omega}_{  {\bf k}_{1} {\alpha}_{1}  } +
{\Omega}_{  {\bf k}_{2} {\alpha}_{2}  }
\right)
\right]
f({\bf k}_{1}, {\alpha}_{1}; {\bf k}_{2}, {\alpha}_{2}|t)
 =
\frac{1}{2V} \sum_{  {\bf k}_{3},  {\bf k}_{4}   }  \>
\sum_{   {\alpha}_{3},   {\alpha}_{4} = 1,2    }
\delta_{   {\bf k}_{1} + {\bf k}_{2},  {\bf k}_{3} + {\bf k}_{4}   }
$$
$$
\times
\left[
{\nu}_{\, {\bf k}_{1} - {\bf k}_{4}  }
\left(
{\cal P}_{  {\bf k}_{1} {\bf k}_{3}  }\right)_{  {\alpha}_{1}  {\alpha}_{3} }
\left(
{\cal P}_{  {\bf k}_{2} {\bf k}_{4}  }\right)_{  {\alpha}_{2}  {\alpha}_{4} }
  +
{\nu}_{\,{\bf k}_{1} - {\bf k}_{3}}
\left(
{\cal P}_{  {\bf k}_{1}  {\bf k}_{4}  }\right)_{  {\alpha}_{1}  {\alpha}_{4} }
\left(
{\cal P}_{  {\bf k}_{2}  {\bf k}_{3}  }\right)_{  {\alpha}_{2}  {\alpha}_{3} }
\right]
$$
$$
\times
\left[
f({\bf k}_{3}, {\alpha}_{3}; {\bf k}_{4}, {\alpha}_{4}|t)
 +
{\Psi }_{  {\bf k}_{3}  {\alpha}_{3}  } (t)
{\Psi }_{   {\bf k}_{4}  {\alpha}_{4}  }(t)
\right]
+
\frac{1}{V}\sum_{  {\bf k}_{3}, {\bf k}_{4}, {\bf k}_{5}  }\>
\sum_{{\alpha}_{3}, {\alpha}_{4}, {\alpha}_{5} = 1,2}
\delta_{  {\bf k}_{1} + {\bf k}_{3},{\bf k}_{4} + {\bf k}_{5}  }
$$
$$
\times
\left[
{\nu}_{\,{\bf k}_{1} - {\bf k}_{5}}
\left(
{\cal P}_{  {\bf k}_{1} {\bf k}_{4}  }\right)_{ {\alpha}_{1} {\alpha}_{4} }
\left(
{\cal P}_{  {\bf k}_{3} {\bf k}_{5}  }\right)_{ {\alpha}_{3} {\alpha}_{5} }
  +
{\nu}_{\,{\bf k}_{1} - {\bf k}_{4}}
\left(
{\cal P}_{  {\bf k}_{1}  {\bf k}_{5}  }\right)_{{\alpha}_{1} {\alpha}_{5}}
\left(
{\cal P}_{  {\bf k}_{3}  {\bf k}_{4}  }\right)_{{\alpha}_{3},{\alpha}_{4}}
\right]
$$
$$
\times
\biggl\{
f({\bf k}_{2}, {\alpha}_{2}; {\bf k}_{5}, {\alpha}_{5}|t)
\left[
n({\bf k}_{3}, {\alpha}_{3}; {\bf k}_{4}, {\alpha}_{4}|t)
+
{\Psi }_{  {\bf k}_{3}  {\alpha}_{3}  }^{\>\ast} (t)
{\Psi }_{   {\bf k}_{4}  {\alpha}_{4}  }(t)
\right]
$$
\begin{equation}
 +
\frac{1}{2}
n^{\>\ast}({\bf k}_{2}, {\alpha}_{2}; {\bf k}_{3}, {\alpha}_{3}|t)
\left[
f({\bf k}_{5}, {\alpha}_{5}; {\bf k}_{4}, {\alpha}_{4}|t)
+
{\Psi }_{  {\bf k}_{5}  {\alpha}_{5}  } (t)
{\Psi }_{   {\bf k}_{4}  {\alpha}_{4}  }(t)
\right]
\biggr\}
 +
\mbox{\rm Idem}
\left[
\left(
{\bf k}_{1}{\alpha}_{1}
\right)
\leftrightarrow
\left(
{\bf k}_{2}{\alpha}_{2}
\right)
\right]\>.  \label{eq_66}
\end{equation}
Here
\begin{equation}
\begin{array}{ccccc}
n({\bf k}_{1}, {\alpha}_{1}; {\bf k}_{2}, {\alpha}_{2}|t) & = &
n^{\>\ast} ({\bf k}_{2}, {\alpha}_{2}; {\bf k}_{1}, {\alpha}_{1}|t) & = &
\langle {\Phi}^{\>+}_{{\bf k}_{1} {\alpha}_{1}}
{\Phi}_{{\bf k}_{2} {\alpha}_{2}}\rangle _{t} -
{\Psi}^{\>\ast}_{{\bf k}_{1} {\alpha}_{1}}(t)
{\Psi}_{{\bf k}_{2} {\alpha}_{2}}(t)\>, \\
 &  &  &  &  \nonumber \\
f({\bf k}_{1}, {\alpha}_{1}; {\bf k}_{2}, {\alpha}_{2}|t) & = &
f ({\bf k}_{2}, {\alpha}_{2}; {\bf k}_{1}, {\alpha}_{1}|t) & = &
\langle{\Phi}_{{\bf k}_{1} {\alpha}_{1}}
{\Phi}_{{\bf k}_{2} {\alpha}_{2}}\rangle _{t} -
{\Psi}_{{\bf k}_{1} {\alpha}_{1}}(t)
{\Psi}_{{\bf k}_{2} {\alpha}_{2}} (t)\>.   \label{eq_35}
\end{array}
\end{equation}

     Equations~(\ref{eq_64})-(\ref{eq_66}) can be reduced to a simpler form
in the physical situation when an external classical source is switched on
and the system is prepared so that at $t = t_{0}$  only one macroscopically
occupied coherent mode ${\bf k} = {\bf k}_{0}$ on the lower polariton
branch is excited:
\begin{equation}
{\Psi}_{{\bf k} \alpha} (t_{0})
\propto
\sqrt{V}
{\delta}_{\alpha , 1}
{\delta}_{{\bf k}, {\bf k}_{0}}\>,\quad
n({\bf k}_{1}, {\alpha}_{1}; {\bf k}_{2}, {\alpha}_{2}|t_{0})
=
f({\bf k}_{1}, {\alpha}_{1}; {\bf k}_{2}, {\alpha}_{2}|t_{0})
=
0\>.    \label{eq_67}
\end{equation}

     From~(\ref{eq_64})-(\ref{eq_66}) we can find the variation of the
functions ${\Psi}_{{\bf k} \alpha} (t)$  and~(\ref{eq_35}) at the time
moment $t = t_{0} + dt$  :
$$
d{\Psi}_{{\bf k} \alpha}
\propto
{\delta}_{{\bf k}, {\bf k}_{0}}\>,\quad
dn({\bf k}_{1}, {\alpha}_{1}; {\bf k}_{2}, {\alpha}_{2})
=
0\>,\quad
df({\bf k}_{1}, {\alpha}_{1}; {\bf k}_{2}, {\alpha}_{2})
\propto
{\delta}_{{\bf k}_{1} + {\bf k}_{2}, 2{\bf k}_{0}}\>.
$$
Therefore, the coherent polariton wave of macroscopic amplitude is
a source of quantum fluctuations, described by the function
$f({\bf k}_{1}, {\alpha}_{1}; 2{\bf k}_{0} - {\bf k}_{1}, {\alpha}_{2}|t)$.
Substituting
$$
{\Psi}_{{\bf k} \alpha} (t)
=
{\delta}_{{\bf k}, {\bf k}_{0}}
{\Psi}_{{\bf k}_{0} \alpha} (t)\>,\quad
n({\bf k}_{1}, {\alpha}_{1}; {\bf k}_{2}, {\alpha}_{2}|t)
=
0\>,
$$
$$
f({\bf k}_{1}, {\alpha}_{1}; {\bf k}_{2}, {\alpha}_{2}|t)
=
{\delta}_{{\bf k}_{1} + {\bf k}_{2}, 2{\bf k}_{0}}\>
f({\bf k}_{1}, {\alpha}_{1}; 2{\bf k}_{0} - {\bf k}_{1}, {\alpha}_{2}|t)
$$
into the right-hand sides of~(\ref{eq_64})-(\ref{eq_66}) we obtain
$$
d{\Psi}_{{\bf k} \alpha}
\propto
{\delta}_{{\bf k}, {\bf k}_{0}}\>,\quad
dn({\bf k}_{1}, {\alpha}_{1}; {\bf k}_{2}, {\alpha}_{2}|t)
\propto
{\delta}_{{\bf k}_{1}, {\bf k}_{2}}\>,\quad
df({\bf k}_{1}, {\alpha}_{1}; {\bf k}_{2}, {\alpha}_{2}|t)
\propto
{\delta}_{{\bf k}_{1} + {\bf k}_{2}, 2{\bf k}_{0}}\>.
$$
This means that after the quantum fluctuations, described by  abnormal
distribution function
$f({\bf k}_{1}, {\alpha}_{1}; 2{\bf k}_{0} - {\bf k}_{1}, {\alpha}_{2}|t)$,
fluctuations  that are characterized by normal (usual) distribution
function  $n({\bf k}_{1}, {\alpha}_{1}; {\bf k}_{1}, {\alpha}_{2}|t)$ are
excited.

      Substituting
$$
{\Psi}_{{\bf k} \alpha} (t)
=
{\delta}_{{\bf k}, {\bf k}_{0}}
{\Psi}_{{\bf k}_{0} \alpha} (t)\>,\quad
n({\bf k}_{1}, {\alpha}_{1}; {\bf k}_{2}, {\alpha}_{2}|t)
=
{\delta}_{{\bf k}_{1}, {\bf k}_{2}}
n({\bf k}_{1}, {\alpha}_{1}; {\bf k}_{1}, {\alpha}_{2}|t)\>,
$$
\begin{equation}
f({\bf k}_{1}, {\alpha}_{1}; {\bf k}_{2}, {\alpha}_{2}|t)
=
{\delta}_{{\bf k}_{1} + {\bf k}_{2}, 2{\bf k}_{0}}
f({\bf k}_{1}, {\alpha}_{1}; 2{\bf k}_{0} - {\bf k}_{1}, {\alpha}_{2}|t)
\label{eq_68}
\end{equation}
into the right-hand sides of~(\ref{eq_64})-(\ref{eq_66}) we can easily
see that the left-hand sides of these equations have the same structure.
Therefore, solution to the~(\ref{eq_64})-(\ref{eq_66}) under the initial
conditions~(\ref{eq_67}) has the form~(\ref{eq_68}). This result is nothing
but a consequence of the momentum
conservation law in the elementary acts of polariton scattering.
     Further simplification of~(\ref{eq_64})-(\ref{eq_66}) can be achieved
by retaining in their right-hand sides only the resonant terms corresponding
to lower polariton branch. Taking into account~(\ref{eq_68}) and
omitting fast-oscillating terms in~(\ref{eq_64})-(\ref{eq_66}) we obtain
\begin{eqnarray}
i\hbar \frac{d}{dt} {\Psi}_{{\bf k}_{0}}(t) & = & [\hbar\Omega_{{\bf k}_{0}}+
{\tilde {\cal F}}_{{\bf k}_{0}}(t)]{\Psi}_{{\bf k}_{0}}(t) + \biggl[{\nu}_{{\mbox{\rm pol}}}({\bf k}_{0},
{\bf k}_{0})\frac{{\Psi}_{{\bf k}_{0}}^2(t)}{V} + {\cal F}_{{\bf k}_{0}}(t)\biggr]
{{\Psi_{{\bf k}_{0}}^{\ast}}}(t)\>;\label{e:psi} \\
 &  &  \nonumber \\
i\hbar \frac{d}{dt} {n_{\bf k}}(t) & = & f_{\bf k}^{\ast}(t)\biggl[{\nu}_{{\mbox{\rm pol}}}({\bf k},
{\bf k}_{0})\frac{{\Psi}_{{\bf k}_{0}}^2(t)}
{V} + {\cal F}_{\bf k}(t)\biggr] - f_{\bf k}(t)\biggl[{\nu}_{{\mbox{\rm pol}}}({\bf k},
{\bf k}_{0})\frac{{\Psi}_{{\bf k}_{0}}^{{\ast}2}(t)}
{V} + {\cal F}_{\bf k}^{\,\ast}(t)\biggr]\>;\label{e:en}   \\
 &  &  \nonumber \\
i\hbar \frac{d}{dt} {f_{\bf k}}(t) & = & \biggl\{ \biggl[\hbar\Omega_{\bf k}+
{\tilde {\nu}}_{{\mbox{\rm pol}}}({\bf k},{\bf k}_{0})\frac{|{\Psi}_{{\bf k}_{0}}(t)|^2}{V} +
 {\tilde {\cal F}}_{\bf k}(t)\biggr]  \nonumber \\
& + & \biggl[\hbar\Omega_{2{\bf k}_{0}-{\bf k}}+{\tilde{\nu}}_{{\mbox{\rm pol}}}(2{\bf k}_{0}-{\bf k},
{\bf k}_{0})\frac{|{\Psi}_{{\bf k}_{0}}(t)|^2}{V} +
{\tilde{\cal F}}_{2{\bf k}_{0}-{\bf k}}(t)\biggr]\biggr\}{f_{\bf k}}(t)   \nonumber \\
& + & [1+{n_{\bf k}}(t)+{n_{2{\bf k}_{0}-{\bf k}}}(t)]\cdot
\biggl[{\nu}_{{\mbox{\rm pol}}}({\bf k},{\bf k}_{0})\frac{{\Psi}_{{\bf k}_{0}}^2(t)}{V} +
{\cal F}_{\bf k}(t)\biggr].\label{e:ef}
\end{eqnarray}
Here
$$
{\cal F}_{\bf k}(t)   =   {\frac{1}{V}}\sum_{{\bf k}_1}
{\nu}_{{\mbox{\rm pol}}}({\bf k},{\bf k}_{1}) {f_{{\bf k}_1}}(t)   =
{\cal F}_{2{\bf k}_{0}-{\bf k}}(t)\>, \qquad
{\tilde {\cal F}}_{\bf k}(t)  =   {\frac{1}{V}}\sum_{{\bf k}_1}
{\tilde {\nu}}_{{\mbox{\rm pol}}}({\bf k},{\bf k}_{1}) {n_{{\bf k}_1}}(t)   =
{\tilde {\cal F}}_{\bf k}^{\,\ast}(t)\>,
$$
and
\begin{eqnarray}
 & {\nu}_{{\mbox{\rm pol}}} ({\bf k}, {\bf k}_{1})  =
 \frac{\displaystyle \mathstrut  {\nu}_{\,   {\bf k} - {\bf k}_{1}  }   }
{\displaystyle
\sqrt
{
\left(1 + L_{\bf k}^{2}\right)
\left(1 + L_{  2{\bf k}_{0} - {\bf k}  }^{2}\right)
\left(1 + L_{  {\bf k}_{1}   }^{2}\right)
\left(1 + L_{  2{\bf k}_{0} - {\bf k}_{1}  }^{2}\right)
}
}\>, &  \nonumber \\
  &    &   \label{pol_const} \\
 & {\tilde \nu}_{{\mbox{\rm pol}}} ({\bf k}, {\bf k}_{1})  =
 \frac{\displaystyle \mathstrut {\nu}_{0} +  {\nu}_{\,   {\bf k} - {\bf k}_{1}  }}
{\displaystyle
\left(1 + L_{\bf k}^{2}\right)
\left(1 + L_{  {\bf k}_{1}   }^{2}\right)
} & \nonumber
\end{eqnarray}
are the polariton-polariton interaction  constants. According
to~(\ref{eq_67}),
the following initial conditions should be imposed on the solution
to~(\ref{e:psi})-(\ref{e:ef}):
\begin{equation}
n_{\bf k}(t_{0}) = f_{\bf k}(t_{0}) = 0\> ,\qquad {\Psi}_{{\bf k}_0}(t_{0})
\neq  0    \>.
\label{in_cond}
\end{equation}
Further we shall assume $t_{0} = 0$.

      In the equilibrium problems  ${\cal F}_{\bf k}$ and
${\tilde {\cal F}}_{\bf k}$ are called the order parameters and are
determined from integral equations \cite{8,Comte}. In the nonequilibrium
problem being studied the order parameters depend on time. Therefore,
their calculation involves  solution of the set of nonlinear
integ\-ro\-dif\-feren\-tial equa\-tions~(\ref{e:psi})-(\ref{e:ef}).

     Equations~(\ref{e:psi})-(\ref{e:ef}) are invariant with respect to time
inversion and do not change under the transformation $t \rightarrow -t$,
${   \Psi_{{\bf k}_{0}}       } \rightarrow
{\Psi_{{\bf k}_{0}}^{\ast}}$, $f_{\bf k} \rightarrow
f_{\bf k}^{\ast}$.

	Besides additive integrals of motion displaying
the conservation laws of the average values of the number of particles,
energy and momentum of a closed system, equations~(\ref{e:psi}) -- (\ref{e:ef})
possess additional integrals of
motion $n_{\bf k}(t)-n_{2{\bf k}_{0}-{\bf k}}(t)=\mbox{\rm const}$ and
${{|f_{\bf k}(t)|}^2}-n_{\bf k}(t)[1+n_{\bf k}(t)]=\mbox{\rm const}$.
Using the initial conditions~(\ref{in_cond}) we obtain
\begin{equation}
n_{\bf k}(t) = n_{2{\bf k}_{0}-{\bf k}}(t) =
\frac{1}{2} \biggl[\sqrt{1+4|f_{\bf k}(t)|^2} - 1\biggr].
\label{link_func}
\end{equation}
Expressions~(\ref{link_func}) enable to exclude equation~(\ref{e:en})
from the set of equations~(\ref{e:psi}) -- (\ref{e:ef}). Note that
representation of type~(\ref{link_func}) were obtained in
\cite{Sc_1986,Sc_1988}.

	It is easy to see that the transformation
$$
f_{\bf k}(t) = \frac{{\Psi}_{{\bf k}_{0}}^2(t)}{|{\Psi}_{{\bf k}_{0}}(t)|^2}
 {g_{\bf k}(t)}
$$
splits out of the set the equation for the phase of the
condensate wave function. As a result, the set of evolution
equations takes the form:
\begin{equation}
\frac{d}{dt}{\cal N}_{0}(t)  =  \frac{2}{\hbar}\, \mbox{\rm Im}\,
{{\cal G}_{{\bf k}_{0}}(t)}
\cdot {\cal N}_{0}(t)\>;\label{e_en}
\end{equation}
\begin{eqnarray}
i\hbar \frac{d}{dt}  {g}_{\bf k}(t) & = &
\biggl\{\hbar \bigl(\Omega_{\bf k}+\Omega_{2{\bf k}_{0}-{\bf k}}-
2\Omega_{{\bf k}_{0}}\bigr) + \left[ {\tilde {\cal F}}_{\bf k}(t) +
 {\tilde{\cal F}}_{2{\bf k}_{0}-{\bf k}}(t) -
 2 {\tilde {\cal F}}_{{\bf k}_{0}}(t)\right]  \nonumber  \\
& + &  [{\tilde \nu}_{{\mbox{\rm pol}}}({\bf k},{\bf k}_{0})+{\tilde \nu}_{{\mbox{\rm pol}}}(2{\bf k}_{0}-{\bf k},{\bf k}_{0})-
2{\nu}_{pol}({\bf k}_{0},{\bf k}_{0})]\frac{|{\Psi}_{{\bf k}_0}(0)|^2}{V}
{\cal N}_{0}(t)-2\mbox{\rm Re}\, {{\cal G}}_{{\bf k}_{0}}(t)\biggr\}
{g_{\bf k}(t)}   \nonumber \\
& + & \sqrt{1+4|g_{\bf k}(t)|^2}\cdot \biggl[{\nu}_{{\mbox{\rm pol}}}({\bf k},{\bf k}_{0})
\frac{|{\Psi}_{{\bf k}_0}(0)|^2}{V}{\cal N}_{0}(t)+
{\cal G}_{\bf k}(t) \biggr] \>,  \label{e_ef}
\end{eqnarray}
where
$$
{{\cal G}}_{\bf k}(t)   =   {\frac{1}{V}}\sum_{{\bf k}_1}
{\nu}_{{\mbox{\rm pol}}}({\bf k},{\bf k}_{1}) {{g}_{{\bf k}_1}}(t)   =
{{\cal G}}_{2{\bf k}_{0}-{\bf k}}(t)  \>,
$$
and
$$
{\cal N}_{0}(t) =  |{\Psi}_{{\bf k}_{0}}(t)|^2
/
 |{\Psi}_{{\bf k}_{0}}(0)|^2
$$
is the relative density of polaritons in the condensate.
%
%

\section{Numerical results\label{num_4.txt}}

	The set of equations~(\ref{e_en}), (\ref{e_ef})  still remains
rather complicated even for numerical methods.
That is why for simplicity we neglect the dispersion
of the exciton-exciton interaction constant $\nu$. In the vicinity of
exciton-photon resonance the function $L_{\bf k} \approx 1$ and,
consequently, ${\tilde\nu}_{{\mbox{\rm pol}}} \approx 2{\nu}_{{\mbox{\rm pol}}}  \approx {\nu}/2$.

	The  initial conditions~(\ref{in_cond}) are the same for all
values of the wave vector $\bf{k}$. Therefore, as follows from
equations~(\ref{e_en}), (\ref{e_ef}),  when there
is no dispersion  of the polariton-polariton
interaction constants, $g$  depends on  $\bf{k}$  via the
functions $\Omega_{\bf k}+\Omega_{2{\bf k}_{0}-{\bf k}}-
2\Omega_{{\bf k}_{0}}$ :
$$
 {g}_{\bf k} \equiv  {g} \bigl(\Omega_{\bf k}+\Omega_{2{\bf k}_{0}-{\bf k}}-
2\Omega_{{\bf k}_{0}}, t\bigr) \>.
$$
Using the identical transformation
$$
   \frac{1}{V}{\sum_{\bf k}}\, {g}\bigl(\Omega_{\bf k}+
\Omega_{2{\bf k}_{0}-{\bf k}}-2\Omega_{{\bf k}_{0}},\,t\bigr)
$$
$$
  = \frac{1}{V}{\sum_{\bf k}}\int\!\!d{x}\>{g}\bigl(\Omega_{\bf k}+
\Omega_{2{\bf k}_{0}-{\bf k}}-2\Omega_{{\bf k}_{0}},\,t\bigr)
{\delta}\bigl(\Omega_{\bf k}+
\Omega_{2{\bf k}_{0}-{\bf k}}-2\Omega_{{\bf k}_{0}}-x\bigr)
$$
$$
 = \int\!\!d{x}\>{g}(x,\,t)\, \frac{1}{V}{\sum_{\bf k}}{\delta}\bigl(\Omega_{\bf k}+
\Omega_{2{\bf k}_{0}-{\bf k}}-2\Omega_{{\bf k}_{0}}-x\bigr)\,,
$$
and moving to the dimensionless variables
\begin{equation}
T=\lambda t\,, \qquad   {w}_{\bf k}=\lambda^{-1}\bigl(\Omega_{\bf k}+
\Omega_{2{\bf k}_{0}-{\bf k}}-2\Omega_{{\bf k}_{0}}\bigr)\,, \qquad
\lambda =
\frac{\nu}{2 \hbar}\,\frac{|{\Psi}_{{\bf k}_0}(0)|^2}{V}\,, \label{dim}
\end{equation}
we obtain the set of equations
\begin{eqnarray}
\frac{d}{d{T}}\, {\cal N}_{0}(T) & = & 2\,\mbox{\rm Im}\,G(T)\cdot
{\cal N}_{0}(T)\,,\label{en_num} \\
 &  &  \nonumber \\
i\frac{\partial}{\partial {T}}\, {g}({w},{T}) & = & [{w} +
{\cal N}_{0}(T)-2\mbox{\rm Re}\,G(T)]{g}({w},{T})  \nonumber \\
 & + & \sqrt{1+4|{g}({w},{T})|^2}
\cdot\biggl[\frac{1}{2}{\cal N}_{0}(T) + G(T)\biggr]\>, \label{ef_num}
\end{eqnarray}
where
\begin{equation}
G(T) = \int\!\!\!d{w}\,\varrho (w){g}({w},{T})\,,
\qquad
\varrho (w) = \frac{1}{2 |{\Psi}_{{\bf k}_0}(0)|^2}\,\sum_{\bf k}
\delta ({w_{\bf k}} - {w})\,. \label{ge}
\end{equation}

   Further the parameters of polaritons formed by mixing of photons and
1A excitons in CdS single crystal are used for numerical estimation:
${\epsilon}_{B}=9.3\,,\> \hbar {\omega}^{\bot}=2.55\,eV\,,\>\eta/\hbar=1.1
\times {10}^{14} \, {c}^{-1} \,, \> m_{\bot}=0.89m_{0}$ ($m_{0}$ is the free
electron mass), and $m_{\|}=2.85m_{0}$ \ \cite{CdS_1}. The  model of isotropic
parabolic exciton band with effective mass $m={(m_{\bot}^{2}m_{\|})}^{1/3}$.
The effective Bohr radius of the 1A exciton in CdS is
$a_{ex}$\ =\ 28~\AA \   \cite{CdS_2}. The exciton ionization energy is
$I_{ex}=27\,meV$ \cite{CdS_3}. The value of the exciton-exciton interaction
constant is evaluated by the formula $\nu \equiv {\nu}_{0}=(26\pi/3)I_{ex}\,
a_{ex}^{3}$ \ \cite{CdS_4} and makes $4.3 \times {10}^{-33}\,erg \cdot
{cm}^{3}$.

	At the initial stage of evolution of the system when
a considerable portion of polaritons is still  in the condensate
and the number of noncondensate polaritons  is rather small we
can omit the terms containing $G(T)$ in  equations~
(\ref{en_num}), (\ref{ef_num}). The solution to the obtained set
of equations has the form
$$
{\cal N}_{0}(T) = \mbox{\rm const} = 1
$$
(the influence of a small portion of noncondensate polaritons on
the condensate is not taken into account), and
\begin{equation}
n({w},{T}) =  \left\{ \begin{array}{ccc}
- \frac{\displaystyle \mathstrut 1}{\displaystyle {w}({w}+2)}
\sinh ^{2}\biggl[\frac{\displaystyle \mathstrut T}{\displaystyle 2}
\sqrt{-{w}({w}+2)}\biggr] & \mbox{\rm for} &
-2 \leq {w} \leq 0 \, \\
 & & \\
\frac{\displaystyle \mathstrut 1}{\displaystyle {w}({w}+2)}
\sin ^{2}\biggl[\frac{\displaystyle \mathstrut T}{\displaystyle 2}
\sqrt{{w}({w}+2)}\biggr] & \mbox{\rm for} &
{w}<-2,\,{w}>0\,.
\end{array}
\right.\label{in_stage}
\end{equation}

    Thus, at the initial stage of evolution the excitation
of noncondensate polaritons takes place in the range $-2\leq w\leq0$
where the distribution function  $n(w)$ has the form of a symmetrical
bell with a maximum at $w=-1$. While moving  away  from the above-mentioned
region the distribution function decreases with oscillations.

The numerical solution of the equation set~(\ref{en_num}), (\ref{ef_num})
was made in two stages. First, the explicit form of the function
$\varrho (w)$ for different values of condensate density
$|{\Psi}_{ {\bf k}_0 }(0)|^2 /V$ and its wave vector ${\bf k}_{0}$
magnitude was found. Further, the function obtained was used to integrate
the evolution equations~(\ref{en_num}), (\ref{ef_num}). The result of
the integration for $|{\bf k}_0| = 3.6 \times
{10}^{5}\,{cm}^{-1}$ are shown in figures 1--3. This choice of
$|{\bf k}_0|$ corresponds to the exciton-photon resonance region located
a bit below the exciton band bottom.

	The decay of the nonequilibrium polariton condensate is
shown in Figure 1. Before the moment of dimensionless time $T
\approx 7$ the condensate depletion  proceeds rather slowly.
Then a drastic fall takes place. As a result, by the moment
$T  \approx  15$  only 10\% of the initial polariton  number remain in
the condensate.
The further condensate decay  is accompanied  by oscillations. This leads to
the  partial restoration of the condensate (up to 30\% at $T \approx 20$).
The oscillations  do not occur  if in
(\ref{ef_num}) the function $G(T)$, describing the integral influence of
all pairs of noncondensate polaritons with the same total momentum
on each individual pair, is neglected. Thus, the occurrence of
oscillations is due to correlation of the states of individual
pairs of noncondensate polaritons. By the moment $T \approx 50$
the oscillations disappear and  further condensate evolution has
the form of slow monotonous decay. These results cannot be obtained by
the introduction of phenomenological constants in the dynamic equations.\\
\parbox[t]{80mm}{
\setlength{\unitlength}{0.240900pt}
\ifx\plotpoint\undefined\newsavebox{\plotpoint}\fi
\sbox{\plotpoint}{\rule[-0.175pt]{0.350pt}{0.350pt}}%
\special{em:linewidth 0.3pt}%
\begin{picture}(825,629)(0,0)
\tenrm
\put(264,158){\special{em:moveto}}
\put(761,158){\special{em:lineto}}
\put(264,158){\special{em:moveto}}
\put(264,516){\special{em:lineto}}
\put(264,158){\special{em:moveto}}
\put(284,158){\special{em:lineto}}
\put(761,158){\special{em:moveto}}
\put(741,158){\special{em:lineto}}
\put(242,158){\makebox(0,0)[r]{0}}
\put(264,194){\special{em:moveto}}
\put(284,194){\special{em:lineto}}
\put(761,194){\special{em:moveto}}
\put(741,194){\special{em:lineto}}
\put(242,194){\makebox(0,0)[r]{0.1}}
\put(264,230){\special{em:moveto}}
\put(284,230){\special{em:lineto}}
\put(761,230){\special{em:moveto}}
\put(741,230){\special{em:lineto}}
\put(242,230){\makebox(0,0)[r]{0.2}}
\put(264,265){\special{em:moveto}}
\put(284,265){\special{em:lineto}}
\put(761,265){\special{em:moveto}}
\put(741,265){\special{em:lineto}}
\put(242,265){\makebox(0,0)[r]{0.3}}
\put(264,301){\special{em:moveto}}
\put(284,301){\special{em:lineto}}
\put(761,301){\special{em:moveto}}
\put(741,301){\special{em:lineto}}
\put(242,301){\makebox(0,0)[r]{0.4}}
\put(264,337){\special{em:moveto}}
\put(284,337){\special{em:lineto}}
\put(761,337){\special{em:moveto}}
\put(741,337){\special{em:lineto}}
\put(242,337){\makebox(0,0)[r]{0.5}}
\put(264,373){\special{em:moveto}}
\put(284,373){\special{em:lineto}}
\put(761,373){\special{em:moveto}}
\put(741,373){\special{em:lineto}}
\put(242,373){\makebox(0,0)[r]{0.6}}
\put(264,409){\special{em:moveto}}
\put(284,409){\special{em:lineto}}
\put(761,409){\special{em:moveto}}
\put(741,409){\special{em:lineto}}
\put(242,409){\makebox(0,0)[r]{0.7}}
\put(264,444){\special{em:moveto}}
\put(284,444){\special{em:lineto}}
\put(761,444){\special{em:moveto}}
\put(741,444){\special{em:lineto}}
\put(242,444){\makebox(0,0)[r]{0.8}}
\put(264,480){\special{em:moveto}}
\put(284,480){\special{em:lineto}}
\put(761,480){\special{em:moveto}}
\put(741,480){\special{em:lineto}}
\put(242,480){\makebox(0,0)[r]{0.9}}
\put(264,516){\special{em:moveto}}
\put(284,516){\special{em:lineto}}
\put(761,516){\special{em:moveto}}
\put(741,516){\special{em:lineto}}
\put(242,516){\makebox(0,0)[r]{1.0}}
\put(264,158){\special{em:moveto}}
\put(264,178){\special{em:lineto}}
\put(264,516){\special{em:moveto}}
\put(264,496){\special{em:lineto}}
\put(264,113){\makebox(0,0){0}}
\put(363,158){\special{em:moveto}}
\put(363,178){\special{em:lineto}}
\put(363,516){\special{em:moveto}}
\put(363,496){\special{em:lineto}}
\put(363,113){\makebox(0,0){20}}
\put(463,158){\special{em:moveto}}
\put(463,178){\special{em:lineto}}
\put(463,516){\special{em:moveto}}
\put(463,496){\special{em:lineto}}
\put(463,113){\makebox(0,0){40}}
\put(562,158){\special{em:moveto}}
\put(562,178){\special{em:lineto}}
\put(562,516){\special{em:moveto}}
\put(562,496){\special{em:lineto}}
\put(562,113){\makebox(0,0){60}}
\put(662,158){\special{em:moveto}}
\put(662,178){\special{em:lineto}}
\put(662,516){\special{em:moveto}}
\put(662,496){\special{em:lineto}}
\put(662,113){\makebox(0,0){80}}
\put(761,158){\special{em:moveto}}
\put(761,178){\special{em:lineto}}
\put(761,516){\special{em:moveto}}
\put(761,496){\special{em:lineto}}
\put(761,113){\makebox(0,0){100}}
\put(264,158){\special{em:moveto}}
\put(761,158){\special{em:lineto}}
\put(761,516){\special{em:lineto}}
\put(264,516){\special{em:lineto}}
\put(264,158){\special{em:lineto}}
\put(686,76){\makebox(0,0)[l]{T}}
\put(214,588){\makebox(0,0)[l]{${\cal N}_{0}$}}
\sbox{\plotpoint}{\rule[-0.500pt]{1.000pt}{1.000pt}}%
\special{em:linewidth 1.0pt}%
\put(269,516){\special{em:moveto}}
\put(274,516){\special{em:lineto}}
\put(279,516){\special{em:lineto}}
\put(284,516){\special{em:lineto}}
\put(289,516){\special{em:lineto}}
\put(294,515){\special{em:lineto}}
\put(299,514){\special{em:lineto}}
\put(304,511){\special{em:lineto}}
\put(309,503){\special{em:lineto}}
\put(314,485){\special{em:lineto}}
\put(319,447){\special{em:lineto}}
\put(324,381){\special{em:lineto}}
\put(329,301){\special{em:lineto}}
\put(334,237){\special{em:lineto}}
\put(339,203){\special{em:lineto}}
\put(344,194){\special{em:lineto}}
\put(348,205){\special{em:lineto}}
\put(353,228){\special{em:lineto}}
\put(358,251){\special{em:lineto}}
\put(363,257){\special{em:lineto}}
\put(368,242){\special{em:lineto}}
\put(373,218){\special{em:lineto}}
\put(378,195){\special{em:lineto}}
\put(383,180){\special{em:lineto}}
\put(388,175){\special{em:lineto}}
\put(393,178){\special{em:lineto}}
\put(398,188){\special{em:lineto}}
\put(403,202){\special{em:lineto}}
\put(408,213){\special{em:lineto}}
\put(413,215){\special{em:lineto}}
\put(418,207){\special{em:lineto}}
\put(423,195){\special{em:lineto}}
\put(428,184){\special{em:lineto}}
\put(433,178){\special{em:lineto}}
\put(438,178){\special{em:lineto}}
\put(443,180){\special{em:lineto}}
\put(448,183){\special{em:lineto}}
\put(453,182){\special{em:lineto}}
\put(458,179){\special{em:lineto}}
\put(463,176){\special{em:lineto}}
\put(468,175){\special{em:lineto}}
\put(473,176){\special{em:lineto}}
\put(478,179){\special{em:lineto}}
\put(483,180){\special{em:lineto}}
\put(488,179){\special{em:lineto}}
\put(493,178){\special{em:lineto}}
\put(497,178){\special{em:lineto}}
\put(502,179){\special{em:lineto}}
\put(507,182){\special{em:lineto}}
\put(512,183){\special{em:lineto}}
\put(517,181){\special{em:lineto}}
\put(522,177){\special{em:lineto}}
\put(527,174){\special{em:lineto}}
\put(532,172){\special{em:lineto}}
\put(537,172){\special{em:lineto}}
\put(542,173){\special{em:lineto}}
\put(547,173){\special{em:lineto}}
\put(552,172){\special{em:lineto}}
\put(557,170){\special{em:lineto}}
\put(562,169){\special{em:lineto}}
\put(567,169){\special{em:lineto}}
\put(572,170){\special{em:lineto}}
\put(577,170){\special{em:lineto}}
\put(582,170){\special{em:lineto}}
\put(587,169){\special{em:lineto}}
\put(592,168){\special{em:lineto}}
\put(597,168){\special{em:lineto}}
\put(602,169){\special{em:lineto}}
\put(607,169){\special{em:lineto}}
\put(612,170){\special{em:lineto}}
\put(617,170){\special{em:lineto}}
\put(622,169){\special{em:lineto}}
\put(627,168){\special{em:lineto}}
\put(632,168){\special{em:lineto}}
\put(637,168){\special{em:lineto}}
\put(642,168){\special{em:lineto}}
\put(647,168){\special{em:lineto}}
\put(651,168){\special{em:lineto}}
\put(656,167){\special{em:lineto}}
\put(661,166){\special{em:lineto}}
\put(666,166){\special{em:lineto}}
\put(671,166){\special{em:lineto}}
\put(676,167){\special{em:lineto}}
\put(681,167){\special{em:lineto}}
\put(686,166){\special{em:lineto}}
\put(691,166){\special{em:lineto}}
\put(696,166){\special{em:lineto}}
\put(701,166){\special{em:lineto}}
\put(706,166){\special{em:lineto}}
\put(711,166){\special{em:lineto}}
\put(716,166){\special{em:lineto}}
\put(721,166){\special{em:lineto}}
\put(726,166){\special{em:lineto}}
\put(731,166){\special{em:lineto}}
\put(736,166){\special{em:lineto}}
\put(741,167){\special{em:lineto}}
\put(746,167){\special{em:lineto}}
\put(751,167){\special{em:lineto}}
\put(756,167){\special{em:lineto}}
\put(761,167){\special{em:lineto}}
\end{picture}
\\[-1.5mm]
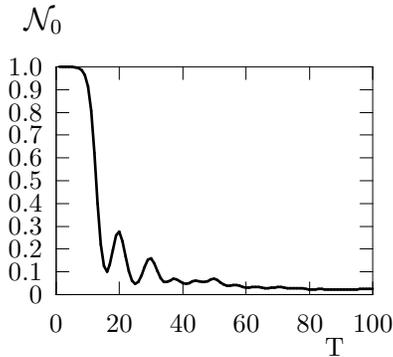
{\footnotesize  Figure~1. Dependence of a relative polariton density in the condensate
on dimensionless time $T$.}
}
\hfill
\parbox[t]{75mm}{
\setlength{\unitlength}{0.240900pt}
\ifx\plotpoint\undefined\newsavebox{\plotpoint}\fi
\sbox{\plotpoint}{\rule[-0.175pt]{0.350pt}{0.350pt}}%
\special{em:linewidth 0.3pt}%
\begin{picture}(825,629)(0,0)
\tenrm
\put(264,158){\special{em:moveto}}
\put(761,158){\special{em:lineto}}
\put(264,158){\special{em:moveto}}
\put(284,158){\special{em:lineto}}
\put(761,158){\special{em:moveto}}
\put(741,158){\special{em:lineto}}
\put(242,158){\makebox(0,0)[r]{0}}
\put(264,209){\special{em:moveto}}
\put(284,209){\special{em:lineto}}
\put(761,209){\special{em:moveto}}
\put(741,209){\special{em:lineto}}
\put(242,209){\makebox(0,0)[r]{0.2}}
\put(264,260){\special{em:moveto}}
\put(284,260){\special{em:lineto}}
\put(761,260){\special{em:moveto}}
\put(741,260){\special{em:lineto}}
\put(242,260){\makebox(0,0)[r]{0.4}}
\put(264,311){\special{em:moveto}}
\put(284,311){\special{em:lineto}}
\put(761,311){\special{em:moveto}}
\put(741,311){\special{em:lineto}}
\put(242,311){\makebox(0,0)[r]{0.6}}
\put(264,363){\special{em:moveto}}
\put(284,363){\special{em:lineto}}
\put(761,363){\special{em:moveto}}
\put(741,363){\special{em:lineto}}
\put(242,363){\makebox(0,0)[r]{0.8}}
\put(264,414){\special{em:moveto}}
\put(284,414){\special{em:lineto}}
\put(761,414){\special{em:moveto}}
\put(741,414){\special{em:lineto}}
\put(242,414){\makebox(0,0)[r]{1.0}}
\put(264,465){\special{em:moveto}}
\put(284,465){\special{em:lineto}}
\put(761,465){\special{em:moveto}}
\put(741,465){\special{em:lineto}}
\put(242,465){\makebox(0,0)[r]{1.2}}
\put(264,516){\special{em:moveto}}
\put(284,516){\special{em:lineto}}
\put(761,516){\special{em:moveto}}
\put(741,516){\special{em:lineto}}
\put(242,516){\makebox(0,0)[r]{1.4}}
\put(264,158){\special{em:moveto}}
\put(264,178){\special{em:lineto}}
\put(264,516){\special{em:moveto}}
\put(264,496){\special{em:lineto}}
\put(264,113){\makebox(0,0){-2}}
\put(347,158){\special{em:moveto}}
\put(347,178){\special{em:lineto}}
\put(347,516){\special{em:moveto}}
\put(347,496){\special{em:lineto}}
\put(347,113){\makebox(0,0){-1.5}}
\put(430,158){\special{em:moveto}}
\put(430,178){\special{em:lineto}}
\put(430,516){\special{em:moveto}}
\put(430,496){\special{em:lineto}}
\put(430,113){\makebox(0,0){-1}}
\put(512,158){\special{em:moveto}}
\put(512,178){\special{em:lineto}}
\put(512,516){\special{em:moveto}}
\put(512,496){\special{em:lineto}}
\put(512,113){\makebox(0,0){-0.5}}
\put(595,158){\special{em:moveto}}
\put(595,178){\special{em:lineto}}
\put(595,516){\special{em:moveto}}
\put(595,496){\special{em:lineto}}
\put(595,113){\makebox(0,0){0}}
\put(678,158){\special{em:moveto}}
\put(678,178){\special{em:lineto}}
\put(678,516){\special{em:moveto}}
\put(678,496){\special{em:lineto}}
\put(678,113){\makebox(0,0){0.5}}
\put(761,158){\special{em:moveto}}
\put(761,178){\special{em:lineto}}
\put(761,516){\special{em:moveto}}
\put(761,496){\special{em:lineto}}
\put(761,113){\makebox(0,0){1}}
\put(264,158){\special{em:moveto}}
\put(761,158){\special{em:lineto}}
\put(761,516){\special{em:lineto}}
\put(264,516){\special{em:lineto}}
\put(264,158){\special{em:lineto}}
\put(711,81){\makebox(0,0)[l]{$w$}}
\put(189,580){\makebox(0,0)[l]{$n(w,T)\times {10}^{-5}$}}
\put(631,451){\makebox(0,0)[r]{{\em T}=10}}
\put(653,451){\special{em:moveto}}
\put(719,451){\special{em:lineto}}
\put(264,158){\special{em:moveto}}
\put(267,158){\special{em:lineto}}
\put(278,158){\special{em:lineto}}
\put(288,159){\special{em:lineto}}
\put(298,159){\special{em:lineto}}
\put(308,160){\special{em:lineto}}
\put(318,161){\special{em:lineto}}
\put(327,163){\special{em:lineto}}
\put(337,164){\special{em:lineto}}
\put(346,166){\special{em:lineto}}
\put(355,168){\special{em:lineto}}
\put(363,171){\special{em:lineto}}
\put(372,173){\special{em:lineto}}
\put(380,175){\special{em:lineto}}
\put(388,177){\special{em:lineto}}
\put(396,179){\special{em:lineto}}
\put(403,181){\special{em:lineto}}
\put(411,182){\special{em:lineto}}
\put(418,183){\special{em:lineto}}
\put(425,183){\special{em:lineto}}
\put(432,184){\special{em:lineto}}
\put(439,183){\special{em:lineto}}
\put(445,183){\special{em:lineto}}
\put(452,182){\special{em:lineto}}
\put(458,181){\special{em:lineto}}
\put(463,180){\special{em:lineto}}
\put(469,179){\special{em:lineto}}
\put(475,177){\special{em:lineto}}
\put(480,176){\special{em:lineto}}
\put(486,175){\special{em:lineto}}
\put(491,173){\special{em:lineto}}
\put(496,172){\special{em:lineto}}
\put(500,171){\special{em:lineto}}
\put(505,169){\special{em:lineto}}
\put(510,168){\special{em:lineto}}
\put(514,167){\special{em:lineto}}
\put(518,166){\special{em:lineto}}
\put(522,165){\special{em:lineto}}
\put(526,164){\special{em:lineto}}
\put(530,164){\special{em:lineto}}
\put(533,163){\special{em:lineto}}
\put(537,162){\special{em:lineto}}
\put(540,162){\special{em:lineto}}
\put(543,161){\special{em:lineto}}
\put(546,161){\special{em:lineto}}
\put(549,161){\special{em:lineto}}
\put(552,160){\special{em:lineto}}
\put(555,160){\special{em:lineto}}
\put(558,160){\special{em:lineto}}
\put(560,160){\special{em:lineto}}
\put(562,159){\special{em:lineto}}
\put(565,159){\special{em:lineto}}
\put(567,159){\special{em:lineto}}
\put(569,159){\special{em:lineto}}
\put(571,159){\special{em:lineto}}
\put(573,159){\special{em:lineto}}
\put(574,159){\special{em:lineto}}
\put(576,159){\special{em:lineto}}
\put(578,159){\special{em:lineto}}
\put(579,159){\special{em:lineto}}
\put(581,158){\special{em:lineto}}
\put(582,158){\special{em:lineto}}
\put(583,158){\special{em:lineto}}
\put(584,158){\special{em:lineto}}
\put(585,158){\special{em:lineto}}
\put(586,158){\special{em:lineto}}
\put(587,158){\special{em:lineto}}
\put(588,158){\special{em:lineto}}
\put(589,158){\special{em:lineto}}
\put(590,158){\special{em:lineto}}
\put(591,158){\special{em:lineto}}
\put(592,158){\special{em:lineto}}
\put(593,158){\special{em:lineto}}
\put(594,158){\special{em:lineto}}
\put(595,158){\special{em:lineto}}
\put(596,158){\special{em:lineto}}
\put(597,158){\special{em:lineto}}
\put(598,158){\special{em:lineto}}
\put(599,158){\special{em:lineto}}
\put(600,158){\special{em:lineto}}
\put(601,158){\special{em:lineto}}
\put(602,158){\special{em:lineto}}
\put(603,158){\special{em:lineto}}
\put(604,158){\special{em:lineto}}
\put(605,158){\special{em:lineto}}
\put(606,158){\special{em:lineto}}
\put(607,158){\special{em:lineto}}
\put(609,158){\special{em:lineto}}
\put(610,158){\special{em:lineto}}
\put(611,158){\special{em:lineto}}
\put(613,158){\special{em:lineto}}
\put(615,158){\special{em:lineto}}
\put(616,158){\special{em:lineto}}
\put(618,158){\special{em:lineto}}
\put(620,158){\special{em:lineto}}
\put(622,158){\special{em:lineto}}
\put(624,158){\special{em:lineto}}
\put(626,158){\special{em:lineto}}
\put(628,158){\special{em:lineto}}
\put(631,158){\special{em:lineto}}
\put(633,158){\special{em:lineto}}
\put(636,158){\special{em:lineto}}
\put(638,158){\special{em:lineto}}
\put(641,158){\special{em:lineto}}
\put(644,158){\special{em:lineto}}
\put(647,158){\special{em:lineto}}
\put(651,158){\special{em:lineto}}
\put(654,158){\special{em:lineto}}
\put(657,158){\special{em:lineto}}
\put(661,158){\special{em:lineto}}
\put(665,158){\special{em:lineto}}
\put(669,158){\special{em:lineto}}
\put(673,158){\special{em:lineto}}
\put(677,158){\special{em:lineto}}
\put(681,158){\special{em:lineto}}
\put(686,158){\special{em:lineto}}
\put(690,158){\special{em:lineto}}
\put(695,158){\special{em:lineto}}
\put(700,158){\special{em:lineto}}
\put(705,158){\special{em:lineto}}
\put(710,158){\special{em:lineto}}
\put(716,158){\special{em:lineto}}
\put(721,158){\special{em:lineto}}
\put(727,158){\special{em:lineto}}
\put(733,158){\special{em:lineto}}
\put(739,158){\special{em:lineto}}
\put(745,158){\special{em:lineto}}
\put(752,158){\special{em:lineto}}
\put(759,158){\special{em:lineto}}
\put(761,158){\special{em:lineto}}
\sbox{\plotpoint}{\rule[-0.500pt]{1.000pt}{1.000pt}}%
\special{em:linewidth 1.0pt}%
\put(631,406){\makebox(0,0)[r]{{\em T}=15}}
\put(653,406){\special{em:moveto}}
\put(719,406){\special{em:lineto}}
\put(264,158){\special{em:moveto}}
\put(267,158){\special{em:lineto}}
\put(278,158){\special{em:lineto}}
\put(288,159){\special{em:lineto}}
\put(298,160){\special{em:lineto}}
\put(308,161){\special{em:lineto}}
\put(318,163){\special{em:lineto}}
\put(327,165){\special{em:lineto}}
\put(337,167){\special{em:lineto}}
\put(346,170){\special{em:lineto}}
\put(355,174){\special{em:lineto}}
\put(363,178){\special{em:lineto}}
\put(372,185){\special{em:lineto}}
\put(380,195){\special{em:lineto}}
\put(388,208){\special{em:lineto}}
\put(396,226){\special{em:lineto}}
\put(403,248){\special{em:lineto}}
\put(411,275){\special{em:lineto}}
\put(418,305){\special{em:lineto}}
\put(425,338){\special{em:lineto}}
\put(432,372){\special{em:lineto}}
\put(439,405){\special{em:lineto}}
\put(445,434){\special{em:lineto}}
\put(452,460){\special{em:lineto}}
\put(458,483){\special{em:lineto}}
\put(463,496){\special{em:lineto}}
\put(469,506){\special{em:lineto}}
\put(475,508){\special{em:lineto}}
\put(480,503){\special{em:lineto}}
\put(486,496){\special{em:lineto}}
\put(491,483){\special{em:lineto}}
\put(496,465){\special{em:lineto}}
\put(500,444){\special{em:lineto}}
\put(505,424){\special{em:lineto}}
\put(510,402){\special{em:lineto}}
\put(514,380){\special{em:lineto}}
\put(518,358){\special{em:lineto}}
\put(522,337){\special{em:lineto}}
\put(526,317){\special{em:lineto}}
\put(530,299){\special{em:lineto}}
\put(533,282){\special{em:lineto}}
\put(537,266){\special{em:lineto}}
\put(540,252){\special{em:lineto}}
\put(543,240){\special{em:lineto}}
\put(546,229){\special{em:lineto}}
\put(549,219){\special{em:lineto}}
\put(552,211){\special{em:lineto}}
\put(555,204){\special{em:lineto}}
\put(558,197){\special{em:lineto}}
\put(560,192){\special{em:lineto}}
\put(562,187){\special{em:lineto}}
\put(565,183){\special{em:lineto}}
\put(567,180){\special{em:lineto}}
\put(569,177){\special{em:lineto}}
\put(571,174){\special{em:lineto}}
\put(573,172){\special{em:lineto}}
\put(574,170){\special{em:lineto}}
\put(576,169){\special{em:lineto}}
\put(578,167){\special{em:lineto}}
\put(579,166){\special{em:lineto}}
\put(581,165){\special{em:lineto}}
\put(582,164){\special{em:lineto}}
\put(583,164){\special{em:lineto}}
\put(584,163){\special{em:lineto}}
\put(585,162){\special{em:lineto}}
\put(586,162){\special{em:lineto}}
\put(587,162){\special{em:lineto}}
\put(588,161){\special{em:lineto}}
\put(589,161){\special{em:lineto}}
\put(590,161){\special{em:lineto}}
\put(591,160){\special{em:lineto}}
\put(592,160){\special{em:lineto}}
\put(593,160){\special{em:lineto}}
\put(594,160){\special{em:lineto}}
\put(595,159){\special{em:lineto}}
\put(596,159){\special{em:lineto}}
\put(597,159){\special{em:lineto}}
\put(598,159){\special{em:lineto}}
\put(599,159){\special{em:lineto}}
\put(600,159){\special{em:lineto}}
\put(601,159){\special{em:lineto}}
\put(602,158){\special{em:lineto}}
\put(603,158){\special{em:lineto}}
\put(604,158){\special{em:lineto}}
\put(605,158){\special{em:lineto}}
\put(606,158){\special{em:lineto}}
\put(607,158){\special{em:lineto}}
\put(609,158){\special{em:lineto}}
\put(610,158){\special{em:lineto}}
\put(611,158){\special{em:lineto}}
\put(613,158){\special{em:lineto}}
\put(615,158){\special{em:lineto}}
\put(616,158){\special{em:lineto}}
\put(618,158){\special{em:lineto}}
\put(620,158){\special{em:lineto}}
\put(622,158){\special{em:lineto}}
\put(624,158){\special{em:lineto}}
\put(626,158){\special{em:lineto}}
\put(628,158){\special{em:lineto}}
\put(631,158){\special{em:lineto}}
\put(633,158){\special{em:lineto}}
\put(636,158){\special{em:lineto}}
\put(638,158){\special{em:lineto}}
\put(641,158){\special{em:lineto}}
\put(644,158){\special{em:lineto}}
\put(647,158){\special{em:lineto}}
\put(651,158){\special{em:lineto}}
\put(654,158){\special{em:lineto}}
\put(657,158){\special{em:lineto}}
\put(661,158){\special{em:lineto}}
\put(665,158){\special{em:lineto}}
\put(669,158){\special{em:lineto}}
\put(673,158){\special{em:lineto}}
\put(677,158){\special{em:lineto}}
\put(681,158){\special{em:lineto}}
\put(686,158){\special{em:lineto}}
\put(690,158){\special{em:lineto}}
\put(695,158){\special{em:lineto}}
\put(700,158){\special{em:lineto}}
\put(705,158){\special{em:lineto}}
\put(710,158){\special{em:lineto}}
\put(716,158){\special{em:lineto}}
\put(721,158){\special{em:lineto}}
\put(727,158){\special{em:lineto}}
\put(733,158){\special{em:lineto}}
\put(739,158){\special{em:lineto}}
\put(745,158){\special{em:lineto}}
\put(752,158){\special{em:lineto}}
\put(759,158){\special{em:lineto}}
\put(761,158){\special{em:lineto}}
\sbox{\plotpoint}{\rule[-0.250pt]{0.500pt}{0.500pt}}%
\special{em:linewidth 0.5pt}%
\put(631,361){\makebox(0,0)[r]{{\em T}=20}}
\put(653,361){\usebox{\plotpoint}}
\put(673,361){\usebox{\plotpoint}}
\put(694,361){\usebox{\plotpoint}}
\put(715,361){\usebox{\plotpoint}}
\put(264,158){\usebox{\plotpoint}}
\put(284,158){\usebox{\plotpoint}}
\put(305,159){\usebox{\plotpoint}}
\put(325,163){\usebox{\plotpoint}}
\put(343,174){\usebox{\plotpoint}}
\put(357,189){\usebox{\plotpoint}}
\put(371,204){\usebox{\plotpoint}}
\put(389,214){\usebox{\plotpoint}}
\put(406,224){\usebox{\plotpoint}}
\put(416,243){\usebox{\plotpoint}}
\put(422,263){\usebox{\plotpoint}}
\put(427,283){\usebox{\plotpoint}}
\put(432,303){\usebox{\plotpoint}}
\put(436,323){\usebox{\plotpoint}}
\put(440,344){\usebox{\plotpoint}}
\put(444,364){\usebox{\plotpoint}}
\put(449,384){\usebox{\plotpoint}}
\put(456,404){\usebox{\plotpoint}}
\put(469,415){\usebox{\plotpoint}}
\put(478,396){\usebox{\plotpoint}}
\put(485,376){\usebox{\plotpoint}}
\put(491,356){\usebox{\plotpoint}}
\put(497,337){\usebox{\plotpoint}}
\put(504,318){\usebox{\plotpoint}}
\put(520,306){\usebox{\plotpoint}}
\put(539,300){\usebox{\plotpoint}}
\put(549,282){\usebox{\plotpoint}}
\put(557,263){\usebox{\plotpoint}}
\put(563,243){\usebox{\plotpoint}}
\put(569,223){\usebox{\plotpoint}}
\put(576,203){\usebox{\plotpoint}}
\put(583,184){\usebox{\plotpoint}}
\put(593,168){\usebox{\plotpoint}}
\put(609,159){\usebox{\plotpoint}}
\put(629,158){\usebox{\plotpoint}}
\put(650,158){\usebox{\plotpoint}}
\put(671,158){\usebox{\plotpoint}}
\put(691,158){\usebox{\plotpoint}}
\put(712,158){\usebox{\plotpoint}}
\put(733,158){\usebox{\plotpoint}}
\put(754,158){\usebox{\plotpoint}}
\end{picture}
\\[-1.5mm]
{\footnotesize  Figure~2. The distribution function  of non-condensate
polaritons at $T =  10, 15, 20$.}
}\\[3.5mm]

      Figure 2  depicts the distribution function of
noncondensate polaritons at time moments in the
interval where the polariton condensate undergoes the most rapid
and significant  changes.
The curve corresponding to the moment of time $T = 10$ as
well as the curve described by the function~(\ref{in_stage}) has the
form of a bell with the maximum at $w = -1$. By the moment $T = 15$ the
distribution function without changing its symmetrical form increases
significantly its value and its maximum shifts to greater energies
of noncondensate polaritons. An increase of the number of
noncondensate polaritons by more than an order of magnitude  corresponds
to the presence of the local minimum of ${\cal N}_{0}(T)$ function  at
$T \approx 15$. When $T = 20$ the distortion of the distribution
function  takes place and the area  under the curve diminishes.
This is in accordance with the local maximum of the ${\cal N}_{0}(T)$
at $T \approx 20$. Thus, the distribution function
is ``breathing'' in conformity with the variation of condensate density. \\[3mm]
\hspace*{15mm}
\parbox[t]{150mm}{
\begin{center}
\hspace*{-15mm}
\setlength{\unitlength}{0.200pt}
\ifx\plotpoint\undefined\newsavebox{\plotpoint}\fi
\sbox{\plotpoint}{\rule[-0.175pt]{0.350pt}{0.350pt}}%
\special{em:linewidth 0.3pt}%

 \\[-12mm]
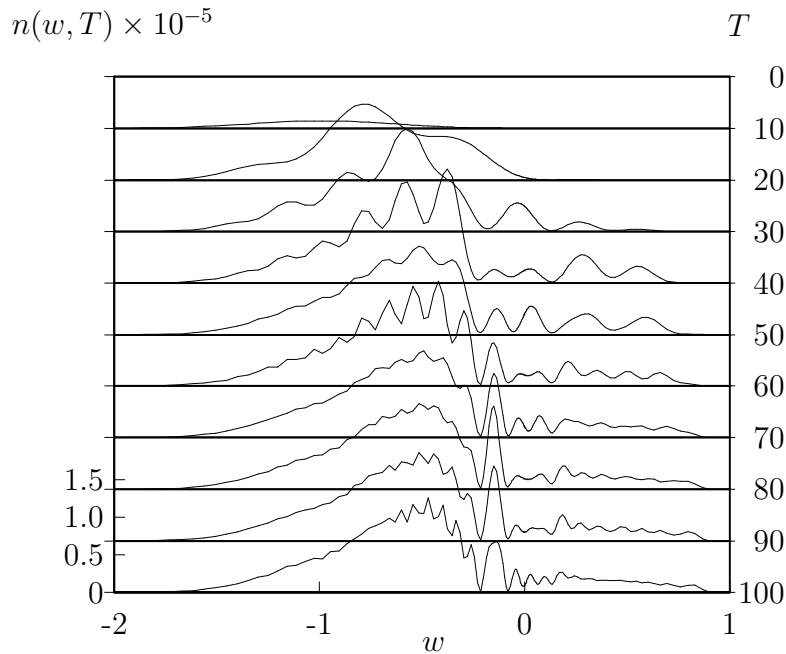
{\footnotesize Figure~3. The  distribution function of non-condensate
polaritons at different moments of dimensionless  time $T$. The
function scale is shown by the marks in the  left vertical axis.}
\end{center}
}\\[4mm]

	Further evolution of the distribution function is
shown in Figure 3. At the moment of time $T = 30$ its contour gets
the oscillating character. The oscillations arise not only in
the dependence of distribution function on parameter $w$ but
in its dependence on dimensionless time $T$  (unfortunately we
could not  demonstrate this in the picture). This means
that if, e.g., the distribution function at the moment
$T \approx 40$ possesses a local maximum at $w = -1$ it can
possess at this point a local minimum at another moment of time.
Later on, the frequency oscillations  become denser, smaller, and
transform into  chaotic ripples  on the smooth
contour of the distribution function.

	From the moment  $T \approx 30$  a short-wave wing begins
to arise in the distribution function curve. By the moment of
time $T = 100$   the function localizes in the interval  $-2 < w < 1$.
Very important is the enhancement and then the partial extinction of the sharp
extra peak localized within the region $-0.22 < w < -0.08$ with a maximum at
$w=-0.15$. The time interval where the extra peak arises and exists is
characterized by small values of condensate density and its relatively
slow change (see Figures 1, 3). The extra peak does not change its position
with time and reaches its maximum at $T \approx 80$.

Under the change of the initial condensate density
$|{\Psi}_{ {\bf k}_0 }(0)|^2 /V$ only the scale of shown dependences changes
due to the corresponding transformation of dimensionless arguments~(\ref{dim}).
This result follows from the very weak dependence of the solution of
equations~(\ref{en_num}), (\ref{ef_num}) on the initial condensate
density via the function $\varrho (w)$ involved in the definitionin of the
$G(T)$-function. According to figures~2,~3, $g(w,T)$ differs from zero
only at $-2 \ < w \ < 1$. The numerical results show that in that region
$\varrho (w)$  does not practically depend on its argument due to
peculiarities of low polariton dispersion in the vicinity of exciton-photon
resonance. This enables us to replace the function $\varrho (w)$ by
$\varrho (0)$ in the definition~(\ref{ge}). As a result we obtain
$G(T) \approx \varrho (0) \int\!\!d{w}\, g(w,T)$. But according to~(\ref{ge})
and~(\ref{dim}) $\varrho (0)$ does not depend on the initial condensate
density and is determined only by the value of wave vector ${\bf k}_{0}$
under the given crystal parameters. The value
$\varrho (0) \sim 6.8 \times {10}^{-6}$ corresponds to the value
$|{\bf k}_0| = 3.6 \times {10}^{5}\,{cm}^{-1}$.
%
%

\section{Conclusions\label{Con}}

      Therefore, the significant condensate depletion takes place already at
the before-kinetic stage of the system evolution. Within a short time
$t_{\ast} \sim 15{\lambda}^{\,-1}$ the condensate density diminishes 10
times compared to the initial density.  For CdS single crystals and
initial condensate density $(10^{\,16} \div  10^{\,18})\, cm^{\,-3}$ the time
$t_{\ast}$ equals to  $(120\, \div 1.2\,)\, ps$. The short time of
the polariton condensate decay makes problematic the possibility of observing
such coherent nonlinear phenomena as, e.g., the phenomenon of self-induced
transparency and creation of solitons in exciton spectral range
\cite{sip_1,sip_2,sip_3,sip_4}.

        Note the irreversible-in-time character of the
described solution of the time-reversible equations \cite{24}.

       We would like to point out also that in \cite{25}
the steady-state  solution
$
|{\Psi}_{{\bf k}_{0}}(t)|^2 = \mbox{\rm const}\,
$,\
$n_{\bf k}(t) \propto
\delta \bigl(E_{\bf k}+
E_{2{\bf k}_{0}-{\bf k}}-2E_{{\bf k}_{0}}\bigr)\,
$,\
$
f_{\bf k}(t) \propto
\delta \bigl(E_{\bf k}+
E_{2{\bf k}_{0}-{\bf k}}-2E_{{\bf k}_{0}}\bigr)\,
$
to equations~(\ref{e:psi}) -- (\ref{e:ef} )
has been obtained (
$
E_{\bf k}
$
is the renormalized energy of the polariton).
Substitution of this solution in the expressions for the
total average energy and total average number of particles shows
that this solution cannot be obtained under the initial
conditions~(\ref{in_cond}).  But the state of the polariton system
corresponding to this solution can be arranged by means of steady laser
action on a semiconductor \cite{26}.
%
%

\section*{Acknowledgments}

We would like to thank Professor~S.~A.~Moskalenko  for stimulating
discussions on the problems of high-density polariton kinetics.
We are also grateful to  Professor~P.~I.~Khadzhi and Dr.~A.~L.~Ivanov
for continued interest in our investigations.
%
%

\end{document}